\documentstyle[twoside,fleqn,espcrc2]{article}
\newcommand{\dintst}{\int \! \! \! \int d^{4}x \, d^{4}y}

\newcommand{\dintsof}{\int_0^{s_0} \! \! \int_0^{s'_0} \!ds\,\; ds'}

\newcommand{\AmS}{{\protect\the\textfont2
  A\kern-.1667em\lower.5ex\hbox{M}\kern-.125emS}}

\title{Vector Meson Dominance, Axial Anomaly
and the Thermal behavior of $g_{\rho \omega \pi}(T)$ }

\author{
{\bf C. A. Dominguez}$^{a}$ and
 {\bf M. Loewe}$^{b}$\\
\vspace{.5cm}
$a$ Institute of Theoretical Physics and Astrophysics,\\ 
    \mbox{} \hspace{.15cm} University of Cape Town, Rondebosch 7700, South 
    Africa\\
$b$ Facultad de Fisica,
    Pontificia Universidad Catolica de Chile, Casilla 306, Santiago 22, Chile}
      
\begin{document}

\begin{abstract}
By using a thermal Finite Energy QCD Sum Rule, we are able to establish the
temperature dependence of the $g_{\omega \rho \pi }(T)$ strong coupling. 
It turns out that this coupling decreases as a function of temperature,
vanishing at the critical temperature. 
This corresponds to a possible deconfining phenomenological signal. This result, 
together with
the Vector Meson Dominance (VMD) expression for the amplitude 
$\pi ^{0} \rightarrow \gamma \gamma $, allows us to establish that
this amplitude also vanishes at the critical temperature, in agreement with previous
independent analysis. This results supports, once again,  the validity
of VMD at finite temperature. Several posssible scenarios
are discussed. However, if VMD would not hold at finite temperature, then
we will not be able to find a prediction for the thermal
behavior of the $\pi ^{0} \rightarrow \gamma \gamma $ 
amplitude.
\end{abstract}

\maketitle

The discussion of the thermal behavior of hadronic Green functions
has called the attention of many authors during the last years, due to their
relevance in the  diagnosis of the Quark-Gluon Plasma  (QGP) \cite{REV1}. Several
correlators have been discussed in the literature, using different techniques
as, for example, QCD (thermal) Sum Rules and different
expansions in the frame of effective low energy hadronic lagrangians.
Among the interesting results, it is worthwhile to mention here that the imaginary part
of all hadronic Green's functions grows with temperature. For a two point Green
function, this means that the hadronic width is an increasing function
with temperature. See \cite{PISA1} and \cite{GAMMAT}. This important result has been confirmed
through many independent analysis. On the other side, it seems that 
there is no such consensus about the temperature dependence of hadronic masses.
Pisarski, for example, \cite{PISA2} has argued that if Vector Meson Dominance
(VMD) is valid at finite temperature, then $M_{\rho }(T_{c}) > M_{\rho}(0)$, being
$T_{c}$ the critical chiral symmetry restoration temperature. Note that $T_{c}$ should
coincide with the critical deconfining temperature $T_{d}$, or if not, both temperatures
should be quite similar, $T_{d} \preceq T_{c}$. If VMD breaks down there is no prediction
for the behavior of $M_{\rho }(T)$. For $T < T_{c}$ different models 
can give opposite answers. Compare, for example,
\cite{PISA1} and \cite{CAD1}. 

\smallskip
The validity of VMD at finite temperature turns out to have a strong impact
on the physics related to the QGP, justifying a careful analysis of this subject
from different
perspectives. In a previous work, we have shown that there is a good agreement between
the prediction for the behavior of the electromagnetic pion form factor $F_{\pi }(T)$,
according to a three-poin function QCD Sum Rule analysis \cite{3PFT}, without
the necessity of invoking VMD, with the VMD expression related to the effective
coupling $g_{\rho \pi \pi}(T)$, which is determined in a completely independent way
\cite{GRHOPI}. The decay $\pi ^{0} \rightarrow \gamma \gamma$ represents another
valuable source of information in order to test the validity of VMD. As it is well known,
the amplitude for this decay, $F_{\pi \gamma \gamma}$, at zero 
temperature and in the chiral limit, 
is determined by the axial $U(1)$ anomaly \cite{ABJ} and can be expressed as
%Eq.1
\begin{equation}
F_{\pi \gamma \gamma} f_\pi = \frac {1}{\pi} \; \alpha_{EM} \; ,
\end{equation}
where $f_\pi \simeq 93$ MeV is the pion decay constant. 
In the previous expression the right side comes from the anomaly.
It has been shown that the anomaly
does not change due to thermal corrections \cite{ANT}. We could then 
naively think that the product
$F_{\pi \gamma \gamma} f_\pi$ will be temperature independent
which, in its turn, would imply that the amplitude $F_{\pi \gamma \gamma}$ should
diverge at the critical temperature. However, as it was disussed in \cite{PISA3},
this statement is not true since the relation between the decay amplitude
and the anomaly does not hold at finite temperature. This is related
to the breaking of the Lorentz invariance at finite temperature. In fact, it can be shown
that $F_{\pi \gamma \gamma}$ vanishes like $f_{\pi }(T)$ at the critical temperature.
Note that this naive scenario, if we assume the validity of VMD, would also imply
the divergence of the effective strong coupling $g_{\rho \omega \pi}$
at $T=T_{c}$, contrary to our expectations about thermal deconfinement.

\smallskip
In this paper, we will find the thermal evolution of $g_{\rho \omega \pi}(T)$
using a thermal Finite Energy QCD Sum Rule (FESR). More precisely, this FESR
will fix the relation between $g_{\rho \omega \pi}(T)$ and the photon-vector meson
coupling. Then, assuming the validity of VMD, we will be able to find 
$F_{\pi \gamma \gamma}(T)$. This analysis confirms the result found in \cite{PISA3},
and suggests, once again, the validity of VMD at finite temperature. The specific 
behavior of $F_{\pi \gamma \gamma}(T)$ depends on the behavior of
$f_{\pi }(T)$, which is known from the composite formalism of Barducci et al
\cite{BAR1}, as well as  on $M_{\rho }(T)$ and $M_{\omega }(T)$ which are
model dependent. We will explore different possibilities for the thermal 
evolution of these masses and compare our result for $F_{\pi \gamma \gamma}(T)$
with other determinations. 

\smallskip
Let us start  with the three-point function
%Eq.2
\begin{eqnarray}
\Pi_{\mu \nu} = i^2 \dintst e^{- i (px+qy)}  \nonumber\\
\times \langle 0 \left| T\left(
  J^{(\rho)}_\mu (x)
  J^{(\pi)}_5 (y) J^{(\omega)}_\nu (0)
       \right) \right|0 \rangle  ,
\end{eqnarray}
\noindent
where the currents are given by $J^{(\rho)}_\mu  = :\overline{u} \gamma_\mu  d:$,
$J^{(\pi)}_5  = (m_u+m_d) :\overline{d} i \gamma_5  u:$, and
$J^{(\omega)}_\nu  = \frac{1}{6} :(\overline{u} \gamma_\nu  u
+\overline{d} \gamma_\nu  d):$, $q = p' - p$,
and the following Lorentz decomposition will be used
%Eq.3
\begin{equation}
  \Pi_{\mu \nu} (p,p',q) = \epsilon_{\mu \nu \alpha \beta} \; p^{\alpha}
  p'^{\beta} \; \Pi (p^2,p'^2,q^2).
\end{equation}

\smallskip
It is easy to convince ourselves that the perturbative contribution
to our three point function, to leading
order in the strong coupling and in the quark masses, 
vanishes identically since it involves
$ {\it Tr} (\gamma_5 \gamma_\alpha
\gamma_\beta \gamma_\rho \gamma_\sigma \gamma_\tau) \equiv 0$. The 
non vanishing perturbative
contribution turns out to be of the order $m_q^2$ and can be
safely neglected. The gluon condensate contribution also vanishes
due to the same argument. So, this leaves us with the leading 
non-perturbative contribution associated with the quark condensate
%Eq.4
\begin{eqnarray}
\Pi(p^2,p'^2,q^2)|_{\mbox{QCD}} = \frac{1}{6} (m_u+m_d) \left(\langle
\overline{u}
u \rangle + \langle \overline{d} d \rangle \right)\nonumber\\
\times \left( \frac{1}{p^2 p'^2}+ \frac{1}{p^2 q^2}+ \frac{1}{p'^2 q^2}
\right)\;,
\end{eqnarray}
\noindent
where $p^{2}$ and $p'^2$ lie in the deep euclidean region, and $q^2$ is
fixed and arbitrary. The isospin $SU(2)$ vacuum symmetry will be assumed in
what follows, i.e. $\langle \overline{u} u \rangle \simeq \langle \overline{d} d \rangle
\equiv  \langle \overline{q} q \rangle$. This expression coincides
with the results of \cite{PAVER} once we convert to their kinematics.

\smallskip
The hadronic representation of our three point correlator can be
constructed inserting one rho- and one omega- intermediate states
using the following definitions
%Eq.5
\begin{equation}
\langle 0 \left | J^{\rho}_\mu \right | \rho^+ \rangle
= \sqrt{2} \; \; \frac{M^2_\rho}{f_\rho} \;\epsilon_\mu \;,
\end{equation}
%Eq.6
\begin{equation}
\langle 0 \left | J^{\omega}_\mu \right | \omega \rangle
= \frac{M^2_\omega}{f_\omega} \; \epsilon_\mu \;,
\end{equation}
%Eq.7
\begin{eqnarray}
\langle \rho (k_1,\epsilon_1) \pi (q) | \omega (k_2,\epsilon_2)
\rangle
= g_{\omega \rho \pi} \epsilon _{\mu \nu \alpha \beta} \nonumber\\
\times \epsilon_1 ^ {\mu} \epsilon_2 ^ {\nu} k_1^{\alpha} k_2^{\beta},
\end{eqnarray}
\noindent
being $\epsilon_1 ^ {\mu}$ the polarization four vectors. 
In this way, and working in the chiral limit, we obtain
%Eq.8
\begin{eqnarray}
\Pi(p^2,p'^2,q^2) |_{\mbox{HAD}}= 2 \; \frac{M^2_{\rho}}{f_\rho}
\frac{M^2_{\omega}}{f_\omega} \frac{f_\pi \mu_\pi^2}{q^2}\nonumber\\
\times \frac{g_{\omega \rho \pi}}{(p^2-M^2_\rho)(p'^2-M^2_\omega)},
\end{eqnarray}

\smallskip
It is interesting to remark that the chiral limit approximation
for the pion propagator
is consistent with having used massless quark propagators in our QCD
calculation. The term $f_{\pi }\mu _{\pi }^{2}$ in Eq.(8), on account
of the Gell-Mann, Oakes and Renner (GMOR) relation, can be written as
$(m_{u} +m_{d})\langle \overline{q} q \rangle /f_{\pi}$ 

\smallskip
Assuming global quark-hadron duality, we are able
to establish the following lowest dimensional FESR for $g_{\omega \rho \pi}$
%Eq.9
\begin{eqnarray}
\dintsof \;Im \; \Pi(s,s',q^2)_{\mbox{HAD}} =\nonumber\\
\dintsof \;Im \; \Pi(s,s',q^2)_{\mbox{QCD}} ,
\end{eqnarray}
where $s=p^{2}$, $s'=p'^{2}$, and $s=s_{0}$ and $s'=s'_{0}$, are the usual
continuum  thresholds. Note that we are dispersing in the $\omega $ and $\rho $ legs.
From this FESR, and in the limit where $q^{2} \rightarrow 0$ we obtain the relation
%Eq.10
\begin{equation}
g_{\omega \rho \pi} = \frac{1}{6} \frac{f_\rho}{M^2_\rho}
\frac{f_\omega}{M^2_\omega} 
\frac{\left [ - (m_u + m_d)
\langle \overline{q}q \rangle \right]}{f_\pi \mu^2_\pi} \left(s_0 + s'_0
\right)
\end{equation}

\smallskip
It is clear that this results does not represent a prediction for $g_{\omega \rho \pi }$,
since we do not know the values of the continuum thresholds $s_{0}$ and $s'_{0}$.
However, since $M_{\rho } \simeq M_{\omega }$, it seems reasonable to set 
$s_{0} = s'_{0}$. If we use the experimental values \cite{PDG}:
$f_\rho = 5.1 \pm 0.3$ and $f_\omega = 15.7 \pm 0.8$,
together with $s_0$ in a  typical range: $\sqrt{s_0} \simeq
1.2 - 1.5$ GeV, Eq. (10) then leads to  $g_{\omega \rho \pi} \simeq
11 - 16 \; \mbox{GeV}^{-1}$, in good agreement with the value extracted
from  $\omega \rightarrow 3 \pi$ decay  ($g_{\omega \rho \pi}  \simeq 16
\; \mbox{GeV}^{-1})$), or the one extracted from 
$\pi^0 \rightarrow \gamma \gamma$
decay using VMD ($g_{\omega \rho \pi} \simeq 11 \; \mbox{GeV}^{-1})$)
\cite{CADG}.

\smallskip
The extension of this analysis to the finite temperature scenario is quite simple.
In principle, all the parameters entering Eq.(10) become temperature dependent.
We also know, from a recent analysis, that there are no temperature corrections
to the GMOR relation at leading order in the quark masses \cite{GMORT}. The temperature
corrections are of the order $m_{q}^2 T^2$, numerically
very small except near the critical temperature for chiral-symmetry
restoration. The temperature dependence
of the continuum threshold $s_{0}(T)$ was first obtained in \cite{DL1} and later
improved in the frame of the composite model \cite{BAR2}. For a wide 
range of temperatures, it turns out from this analysis that the following
scaling relation holds to a good approximation.
%Eq.11
\begin{equation}
\frac{f_{\pi}^{2}(T)}{f_{\pi}^{2}(0)} \simeq
\frac{\langle \bar{q} q\rangle _{T}}{\langle \bar{q} q\rangle _{0}} \simeq
\frac{s_{0}(T)}{s_{0}(0)} \;.
\end{equation}
Hence, Eq. (10) can be recast as
%Eq.12
\begin{eqnarray}
\frac{G(T)}{G(0)} \equiv \frac{g_{\omega\rho\pi}(T)/f_\rho (T) f_\omega (T)}
{g_{\omega\rho\pi}(0)/f_\rho(0) f_\omega(0)}=\nonumber\\
\frac{f_\pi^3(T)}{f_\pi^3(0)}
\frac{1}{M_\rho^2(T)/M_\rho^2(0)}
\frac{1}{M_\omega^2(T)/M_\omega^2(0)}.
\end{eqnarray}
\smallskip
Note that the function $G(T)$ defined above  is the VMD expression for the 
$\pi ^{0} \rightarrow \gamma \gamma$  amplitude $F_{\pi \gamma \gamma}$ of
Eq.(1), viz.
%Eq.13
\begin{equation}
F_{\pi\gamma\gamma}|_{VMD} = 8 \pi \alpha_{EM} \frac{g_{\omega\rho\pi}}
{f_\rho f_\omega} \; ,
\end{equation}
so that
%Eq.14
\begin{equation}
\frac{F_{\pi\gamma\gamma}(T)}{F_{\pi\gamma\gamma}(0)}|_{VMD}
= \frac{G(T)}{G(0)} \;.
\end{equation}

\smallskip
While $f_{\pi }(T)$ is known \cite{BAR1}, the thermal behavior of the vector meson
masses is still a matter of controversy. Let us consider the behavior of $G(T)$
according to different possibilities for the thermal vector meson masses: 
(a) If $M_\rho(T) \simeq M_\rho(0)$ and
$M_\omega(T) \simeq M_\omega(0)$ then $G(T)$ vanishes as $f_\pi^3(T)$
as $T \rightarrow T_c$;

\noindent
(b) If $M_\rho(T) > M_\rho(0)$ and
$M_\omega(T) \simeq M_\omega(0)$ then $G(T)$ still vanishes as $f_\pi^3(T)$. This
possibility is a consequence of VMD \cite{PISA2}. It is nice to see that in this
case $F_{\pi \gamma \gamma }(T_{c})|_{VMD} = 0$. The vanishing of the decay amplitude
at the critical temperature has been confirmed in the frame
pf some field theory model \cite{PISA3};

\noindent
(c) If both $M_\rho(T)$ and $M_\omega(T)$ vanish at $T=T_c$ as
$f_\pi(T)$, then
$G(T)$ diverges as $1/f_\pi(T)$; this possibility is just a trivial property
of the bag model, where everything scales as $f_\pi(T)$. Note that this behavior does
not contradict the expectation that $F_{\pi \gamma \gamma }(T_{c}) = 0$, since
such kind of thermal behavior for the vector masses means that VMD is no longer
valid at finite temperature and, therefore, Eq.(14) is not necessarily true.

\smallskip
\noindent
The dependence of $f_{\rho }(T)$ and $f_{\omega }(T)$ is also needded, in addition to
the thermal vector meson masses, if we are interested in the thermal behavior of
$g_{\omega \rho \pi }(T)$. From our experience 
with other effective hadronic couplings,  as for example 
the thermal behavior of the current-nucleon
coupling \cite{MNT}, $g_{\pi NN}(T)$ \cite{PINUCL}, we could expect
the vanishing of $g_{\omega \rho \pi }(T_{c})$. We  know
that $f_{\rho }(T)$ diminishes with T
as $f_{\rho }(T) =f_{\rho }(1 - T^{2}/12f_{\pi }^{2})$,\cite{LEE}. Note that
in chiral models
$f_\omega$ is temperature independent at leading order because the
omega meson does not couple to two pions. For both possibilities a) and b)
mentioned above, then $g_{\omega \rho \pi }(T)$  would vanish as 
$f_\pi^3(T)$ if
$f_\rho(T) \simeq f_\rho(0)$ and $f_\omega(T) \simeq f_\omega(0)$,
or faster than $f_\pi^3(T)$ if $f_\rho(T_c)=f_\omega(T_c)=0$.
In case (c) $g_{\omega\rho\pi}(T)$
would still vanish as $f_\pi(T)$ because  $f_\rho(T)$ and $f_\omega(T)$
would scale as the vector meson masses. 

\smallskip
In summary, $g_{\omega\rho\pi}(T)$
vanishes in all possible cases, regardless the thermal behavior of the vector
masses. In this sense, it can be considered as a phenomenological order parameter
for the deconfining phase transition.
The results we have presented here can be wieved as supporting
the evidence for VMD at finite temperature.

\medskip
\noindent
\bf Acknowledgements: \rm The work of (CAD) has been supported in part by the FRD
(South Africa), and that of (ML) by Fondecyt (Chile) under grant No. 1980577.

\pagebreak
\noindent
\Large Discussion

\normalsize

\smallskip
\noindent
\bf M. Nielsen, \rm IFUSP, Sao Paulo.

\noindent
\it Is your calculation for $g_{\omega \rho \pi }$ done at $q^{2}=0$?

\smallskip
\noindent
\bf M. Loewe

\noindent
\it Yes. It is a non trivial result that we can take this limit safely.

\medskip
\noindent
\bf S. Narison, \rm Montpellier

\noindent
\it What is the thermal behavior of the coupling $g_{\omega \rho \pi}$ in the case where
$M_\rho(T) > M_\rho(0)$ and also $M_\omega(T) > M_\omega(0)$?.

\smallskip
\noindent
\bf M. Loewe

\noindent
\it The coupling would vanish since $F_{\pi \gamma \gamma }(T)$ vanishes like
$f_{\pi }^{3}(T)$, $f_{\rho }$ diminishes with T and $f_{\omega }(T)$ remains
constant. Note, however, that in the gauged linear sigma model discussed by Pisarski
the $M_{\omega }$ does not varied with T.

\medskip
\noindent
\bf M. Eidenm\"{u}ller \rm Heidelberg

\noindent
\it You gave the scaling relation $\frac{\langle \bar{q} q\rangle _{T}}{\langle \bar{q} q\rangle _{0}} \simeq
\frac{s_{0}(T)}{s_{0}(0)}$. How can this be shown? 

\smallskip
\noindent
\bf M. Loewe

\noindent
\it \noindent
\it This results emerges from the determination done by Barducci et al, in the frame of
the composite model, for
the quark condensate at finite temperature. The idea of this approach 
is to introduce the expected
behavior for a second order phase transition near the critical temperature, according
to mean field theory, and, simultaneously,  to reproduce in the low temperature region
the well known results from chiral perturbation theory.

\smallskip
\noindent
We used their result for
the thermal quark condensate as an imput in a sum rule for the correlator
of two axial currents, obtaining $s_{0}(T)$. From this analysis we were able to
establish the 
validity of this scaling relation in a broad range of temperatures.

\end{document}